\documentclass[]{aa}     % LaTeX A&A  Standard Fonts
\usepackage{graphics}
\def\cc{\,{\rm cm^{-3}}}
\def\cm2{\,{\rm cm^{-2}}}
\def\kms{\,{\rm {km\,s^{-1}}}}
\def\kkms{\,{\rm {K\,km s^{-1}}}}
\def\co{\,{\rm ^{12}CO}}
\def\13co{\,{\rm ^{13}CO}}
\def\h2{\,{\rm H_{2}}}
\def\Msun{\rm M_{\odot}}

\def\aua{{\rm A\&A} }

\def\apj{{\rm ApJ} }
\def\aj{{\rm AJ} }
\def\apjs{{\rm ApJS} }
\def\apjl{{\rm ApJL} }

\def\mnras{{\rm MNRAS} }
\def\pasj{{\rm PASJ} }
\def\pasp{{\rm PASP} }
\begin{document}
 
\title{CI and CO in the nearby spiral galaxies IC~342 and Maffei~2}
 
   \subtitle{}
 
\author{F.P. Israel
          \inst{1}
           and F. Baas
          \inst{1,2, \dag}
           }
 
   \offprints{F.P. Israel}
 
  \institute{Sterrewacht Leiden, P.O. Box 9513, 2300 RA Leiden,
             The Netherlands
  \and       Joint Astronomy Centre, 660 N. A'ohoku Pl., Hilo,
             Hawaii, 96720, USA}
 
\date{\dag  Deceased April 4, 2001\\
\\
Received ????; accepted ????}
 
\abstract{
We present $J$=2--1, $J$=3--2, $J$=4--3 $\co$ and 492 GHz [CI] maps 
as well as $J$=2--1 and $J$=3--2 $\13co$ measurements of the central
regions in the nearby Sc galaxies IC~342 and Maffei~2. In both galaxies. 
the distribution of CO and [CI] is strongly concentrated towards the
center. These centers harbour modest starbursts. Both galaxies have 
nearly identical $\co$ transitional ratios
but the relative intensities of their $\13co$ and [CI] emission lines
differ significantly. The observed sets of line intensities require
modelling with a multi-component molecular gas. In both galaxies, a
dense component must be present ($n(\h2) \approx 10^{4} - 10^{5} \cc$) 
with kinetic temperatures $T_{kin}$ = 10--20 K (IC~342) or 20--60 K
(Maffei~2), as well as a less dense (IC~342: a few hundred $\cc$ at
most; Maffei~2: $\approx 3 \times 10^{3} \cc$) and hotter ($T_{kin}$
= 100--150 K) component. In both galaxies, neutral and ionized atomic
carbon amounts are between 1.5 and 2.5 times those of CO.
In both starburst centers about half to two thirds of the molecular 
gas mass is associated with the hot PDR phase. 
The center of IC~342 contains within $R$ = 0.25 kpc an (atomic and 
molecular) gas mass of 1 $\times 10^{7} \Msun$ and a peak face-on 
gas mass density of about 70 $\Msun$\, pc$^{-2}$. For Maffei~2 these
numbers are less clearly defined, mainly because of uncertainties
in its distance and carbon abundance. We find a gass mass $M_{\rm gas}
\geq 0.5 \times 10^{7} \Msun$, and a peak face-on gas mass 
density of about 35 $\Msun$\, pc$^{-2}$.
}

\maketitle
 
\section{Introduction}

%Table 1
\begin{table}
\caption[]{Galaxy parameters}
\begin{flushleft}
\begin{tabular}{lll}
\hline
\noalign{\smallskip}
			    & IC~342				  & Maffei~2 \\
\noalign{\smallskip}
\hline
\noalign{\smallskip}
Type$^{a}$     	    & SABcd				  & SABbc \\
Optical Centre:		    & 					  & \\
R.A. (B1950)$^{b}$ 	    & 03$^{h}$41$^{m}$58.6$^{s}$  	  & 02$^{h}$38$^{m}$08.0$^{s}$ \\
Decl.(B1950)$^{b}$          & +67$^{\circ}$56$'$26$''$ 	          & +59$^{\circ}$23$'$24$''$ \\
R.A. (J2000) 	            & 03$^{h}$46$^{m}$49.7$^{s}$  	  & 02$^{h}$41$^{m}$54.6$^{s}$ \\
Decl.(J2000)                & +68$^{\circ}$05$'$45$''$ 	          & +59$^{\circ}$36$'$11$''$ \\
Radio Centre :		    & \\
R.A. (B1950)$^{c}$ 	    & 03$^{h}$41$^{m}$57.3$^{s}$	  & 02$^{h}$38$^{m}$08.4$^{s}$ \\
Decl.(B1950)$^{c}$           & +67$^{\circ}$56$'$27$''$	  	  & +59$^{\circ}$23$'$30$''$ \\
$V_{\rm LSR}^{c,d}$    	    & +35 $\kms$ 			  & --31 $\kms$ \\
Inclination $i^{d}$ 	    & 25$^{\circ}$ 			  & 67$^{\circ}$ \\
Position angle $P^{d}$      & 39$^{\circ}$ 			  & 26$^{\circ}$ \\
Distance $D^{e}$            & 1.8 Mpc 				  & 2.7 Mpc\\
Luminosity $L_{\rm B}^{f}$  & 2 $\times 10^{10}$ L$_{\rm B\odot}$ & 1 $\times 10^{10}$ L$_{\rm B\odot}$ \\
Scale           	    & 115 $''$/kpc 			  & 77 $''$/kpc \\
\noalign{\smallskip}
\hline
\end{tabular}
\end{flushleft}
Notes to Table 1:\\
$^{a}$ RSA (Sandage $\&$ Tammann 1987) \\
$^{b}$ Dressel $\&$ Condon (1976); \\
$^{c}$ Hummel $\&$ Gr\"ave (1990); Turner $\&$ Ho (1994) \\
$^{d}$ Rots (1979); Newton (1980); Hurt et al. (1996) \\
$^{e}$ McCall (1989) for IC~342; for Maffei~2 see text \\
$^{f}$ Buta $\&$ McCall (1999), assuming Galactic foreground for IC~342 $A_{\rm B}$ = 3.32 mag (Madore $\&$
Freedman (1992); for Maffei~2 an uncertain foreground $A_{\rm B}$ = 8 mag is assumed. 
\end{table}

%Table 2 Log of Observations
\begin{table*}
\caption[]{Observations Log}
\begin{flushleft}
\begin{tabular}{llcccccccccc}
\hline
\noalign{\smallskip}
Transition & Object & Date    	& Freq	& $T_{\rm sys}$ & Beam 	& $\eta _{\rm mb}$ & t(int) & \multicolumn{4}{c}{Map Parameters} \\
& & & & & Size & & & Points & Size & Spacing & P.A. \\
	   &	    & (MM/YY) 	& (GHz)	& (K)	  & ($\arcsec$) & 	       	   & (sec) & & ($\arcsec$) & ($\arcsec$) & ($^{\circ}$)	\\
\noalign{\smallskip}
\hline
\noalign{\smallskip}
$\co$ $J$=2--1	 & IC~342	& 02/89    & 230  & 1290  	& 21	& 0.63  & 600  & 16 & 40$\times$80  &  8 & 70 \\
		 &		& 01/96	   &	  &  640	&	& 0.69  &  60  & 52 & 		    &    &    \\
	   	 & Maffei~2   	& 01/96    &      &  490      	&       & 0.69  & 240  & 25 & 48$\times$60  & 12 & 25 \\
$\co$ $J$=3--2   & IC~342	& 04/94    & 345  &  735  	& 14	& 0.58  & 180  & 39 & 32$\times$72  &  8 & 70 \\
	   	 & Maffei~2	& 12/93    &	  & 1854	& 	& 0.53 	& 400  & 10 & 36$\times$48  &  6 & 27 \\
	   	 &		& 11/94	   & 	  & 1465       	& 	& 0.58  & 240  & 33 & 		    &    &    \\
		 &		& 12/00	   &	  &  565        &       & 0.63  & 120  & 77 & 60$\times$80  &  8 & 27 \\
$\co$ $J$=4--3   & IC~342	& 12/93    & 461  & 8000  	& 11	& 0.53  & 600  &  8 & 24$\times$40  &  8 & 70 \\
		 &		& 04/94    &	  & 2170	&	& 0.53  & 400  & 14 &               &    &    \\
	   	 & Maffei~2	& 04/94    &      & 3200        &	& 0.53  & 480  &  7 & 30$\times$36  &  6 & 25 \\
		 &		& 07/96    &	  & 3700	&       & 0.53	& 360  & 26 &		    &	 &    \\
\noalign{\smallskip}
\hline
\noalign{\smallskip}
%\multicolumn{7}{l}{$^{13}$CO} \\
%\noalign{\smallskip}
$\13co$ $J$=2--1 & IC~342   	& 02-89    & 220   & 1440       & 21    & 0.63 &  900  &  5 & 30$\times$30  & 15 &  0 \\
	   	 & Maffei~2	& 01-96	   &	   &  550	& 	& 0.69 & 2400  &  1 & & & \\
$\13co$ $J$=3--2 & IC~342   	& 04-94    & 330   & 2170       & 14    & 0.58 & 6000  &  1 & & & \\
		 &		& 01-96	   &	   & 1660	&       & 0.58 & 1560  &  5 & 16$\times$16  &  8 &  0 \\
	   	 & Maffei~2	& 01-96    &	   & 2800	&	& 0.58 & 2400  &  1 & & & \\
\noalign{\smallskip}
\hline
\noalign{\smallskip}
%\multicolumn{7}{l}{CI} \\ 
CI $^{3}$P$_{1}$--$^{3}$P$_{0}$ & IC~342 & 11-94   & 492   & 5250	& 10 	& 0.53 &  800  & 27 & 24$\times$48 & 8 & 70 \\
		 & Maffei~2      & 12-93    &       & 4885	&       & 0.53 & 3200  &  1 & & & \\
\noalign{\smallskip}
\hline
\end{tabular}
\end{flushleft}
\end{table*}

Molecular gas is a major constituent of the interstellar medium in galaxies.
Within the inner kiloparsec, many spiral galaxies exhibit a strong 
concentration of molecular gas towards their nucleus. We have conducted a 
programme to observe a number of nearby galaxies in various CO 
transitions and in the 492 GHz $^{3}$P$_{1}$--$^{3}$P$_{0}$ CI transition in 
order to determine the physical condition of such central molecular gas 
concentrations. Results on NGC~253 (Israel, White $\&$ Baas 1995),
NGC~7331 (Israel $\&$ Baas 1999), NGC~6946 and M~83 = NGC~5236 (Israel $\&$ 
Baas 2001) have already been published. In this paper, we present the results 
for the very nearby galaxies IC~342 and Maffei~2 whose basic properties 
are summarized in Table 1.

Both are major members of the IC~342/Maffei group, which is located 
at a distance of about 2 Mpc (Huchtmeier et al., 2000 and references therein).
This appears to be the galaxy group closest to the Local Group but it is, 
unfortunately, located in the sky very close to the Galactic plane. 
Consequently, its members suffer high foreground extinction rendering several 
of them, including Maffei~2, all but invisible at optical wavelengths. 
Mainly for this reason, the distance of Maffei~2 is still quite uncertain.
However, it is usually assumed that it exceeds that of IC~342. We have
thus assumed a distance of 2.7 Mpc which is in line with the overall
cluster value, but 50 per cent greater than the reasonably well established
distance to IC~342. Both IC~342 and Maffei~2 have been studied well at longer 
wavelengths, including those of the CO line. Inasmuch as IC~342 is, together 
with NGC~253 and M~82, one of the strongest (sub)millimeter line emitters in 
the sky, it has been observed frequently. In the $J$=1--0 transition of 
$\co$, it was one of the first galaxies mapped at $\approx
1'$ resolution (Morris $\&$ Lo 1978; Rickard $\&$ Palmer 1981; Young $\&$
Scoville 1982). More extensive mapping of the $\co$ and $\13co$ isotopes
in both the $J$=1--0 and $J$=2--1 transitions was presented by Eckart et al.
(1990) and Xie et al. (1994). Emission in the $J$=3--2 transition was
detected and mapped by Ho et al. (1987), Steppe et al. (1990) and
Irwin $\&$ Avery (1992), while the $\13co$ isotope was detected in
this transition by Wall $\&$ Jaffe (1990) and Mauersberger et al. (1999). 
Finally, the $J$=4--3 and, with a somewhat uncertain calibration, the 
$J$ = 6--5 $\co$ transitions were detected by G\"usten et al. (1993) and 
Harris et al. (1991). The $\co/\13co$ isotope ratio 
was determined to be $\geq 30$ by Henkel et al. (1998). High-resolution 
aperture synthesis maps of IC~342 have been published in $J$=1--0 $\co$ 
and $\13co$ by Ishizuki et al. (1990), Turner $\&$ Hurt (1992), Wright et 
al., (1993) and Sakamoto et al. (1999), and in $J$=2--1$\co$ and $\13co$
by Turner et al., (1993) and Meier et al. (2000). 

Maffei~2 was first detected in $\co$ by Rickard et al. (1977) and 
subsequently mapped at 23$''$ resolution by Weliachew et al. (1988). 
Emission in the $J$=2--1 and $J$=3--2 transitions was measured by Wall et al. 
(1993) and Mauersberger et al. (1999). Maffei~2 was mapped in $J$=1--0 $\co$
and $\13co$ by Ishiguro et al. (1989) and Hurt $\&$ Turner (1991), in $J$=2--1
$\co$ by Sargent et al. (1985) and in $J$=3--2 $\co$ by Hurt et al. (1993).

Although the molecular line emission from Maffei~2 is somewhat weaker than
that from IC~342, both are frequently included in the same observing
programs. Both have been observed in a great variety of other molecular 
species: CS (Mauersberger $\&$ Henkel, 1989; Mauersberger et al., 1989; 
Sage et al., 1990; Paglione et al., 1995), HCN and HCO$^{+}$ (Nguyen-Q-Rieu 
et al., 1992; Downes et al., 1992; Jackson et al. 1995; Paglione et al. 
1997), H$^{13}$CO$^{+}$ and N$_{2}$H$^{+}$ (Mauersberger $\&$ Henkel, 1991); 
HNC (H\"uttemeister et al., 1995), CN and HC$_{3}$N (Henkel et al. (1988), 
HNCO (Nguyen-Q-Rieu et al. 1991), OCS (Mauersberger et al., 1995) as well 
as H$_{2}$CO and CH$_{3}$OH (H\"uttemeister et al., 1997). In particular 
the discovery of strong NH$_{3}$ emission, first from IC~342 (Ho et al. 1990 
and references therein), then from Maffei~2 (Henkel et al. 2000; Takano et 
al. 2000) showed that a significant fraction of the molecular gas in the 
centers of these galaxies is both dense and hot.

%Figure1: Central Profiles
\begin{figure*}[t]
\unitlength1cm
\begin{minipage}[b]{3.94cm}
\resizebox{4.2cm}{!}{\includegraphics*{ic342_co21.ps}}
\end{minipage}
\hfill
\begin{minipage}[t]{3.94cm}
\resizebox{4.2cm}{!}{\includegraphics*{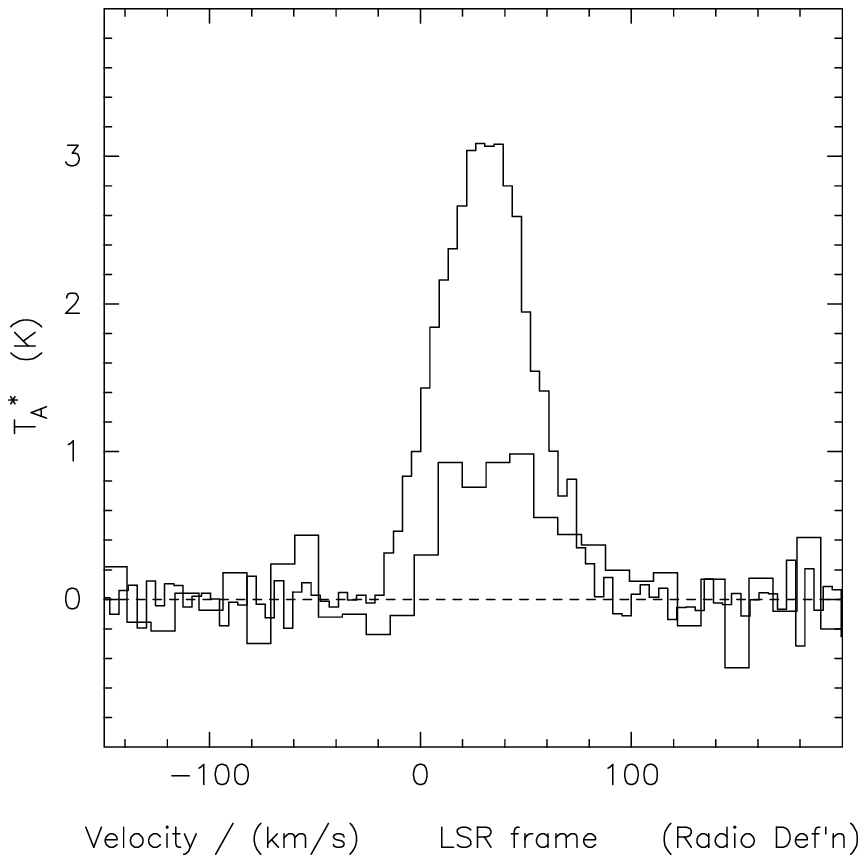}}
\end{minipage}
\hfill
\begin{minipage}[t]{3.94cm}
\resizebox{4.2cm}{!}{\includegraphics*{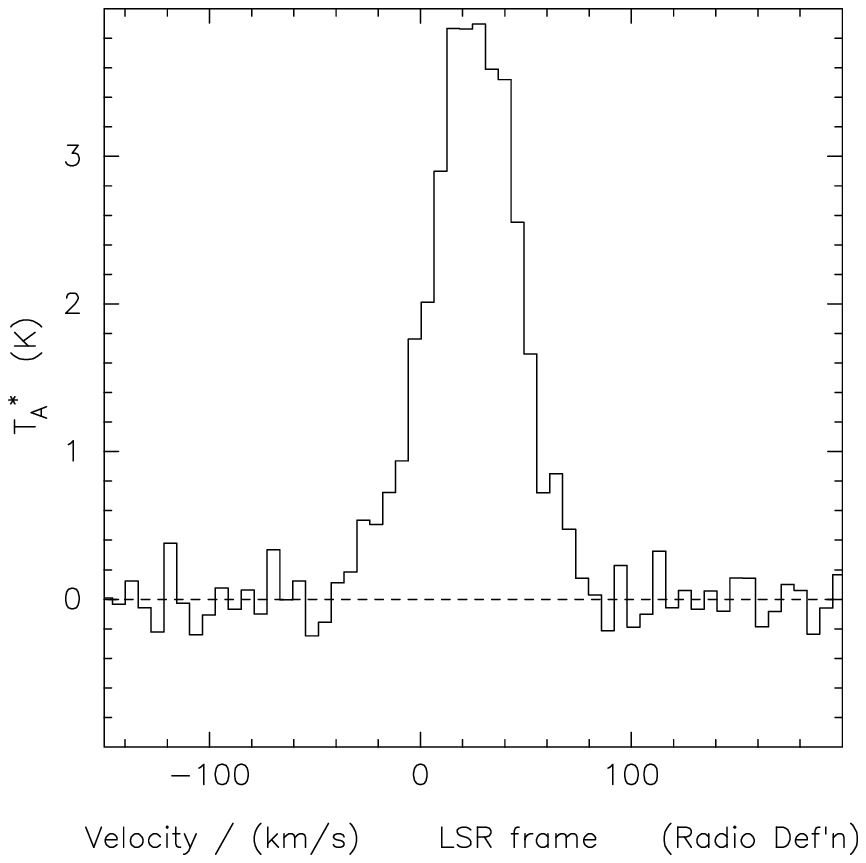}}
\end{minipage}
\hfill
\begin{minipage}[t]{3.94cm}
\resizebox{4.2cm}{!}{\includegraphics*{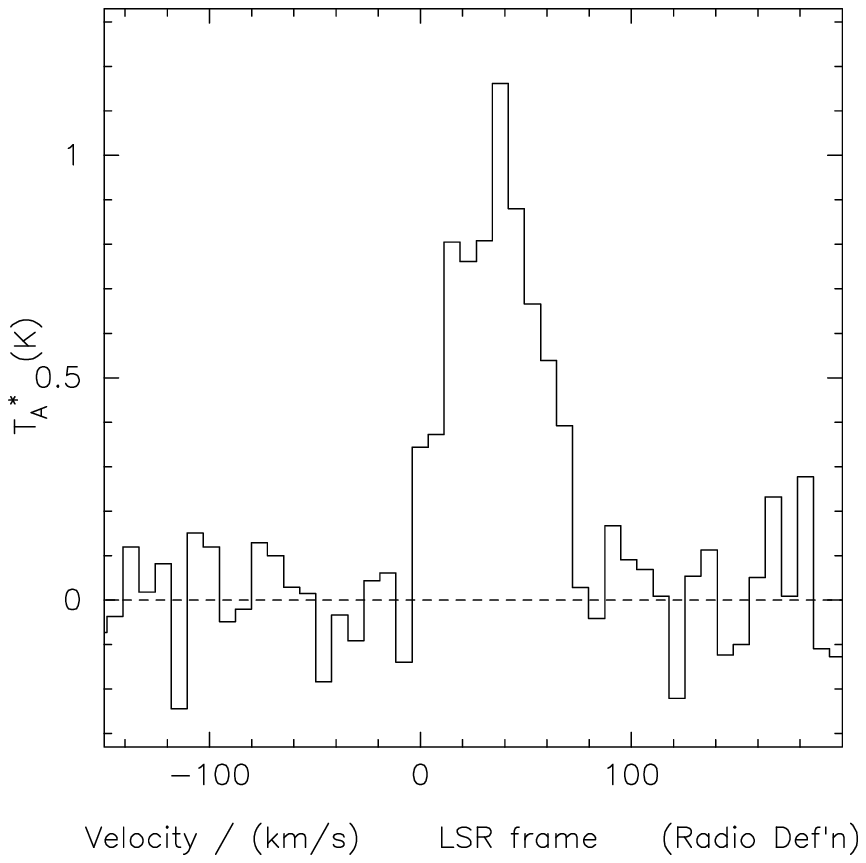}}
\end{minipage}
\begin{minipage}[b]{3.94cm}
\resizebox{4.2cm}{!}{\includegraphics*{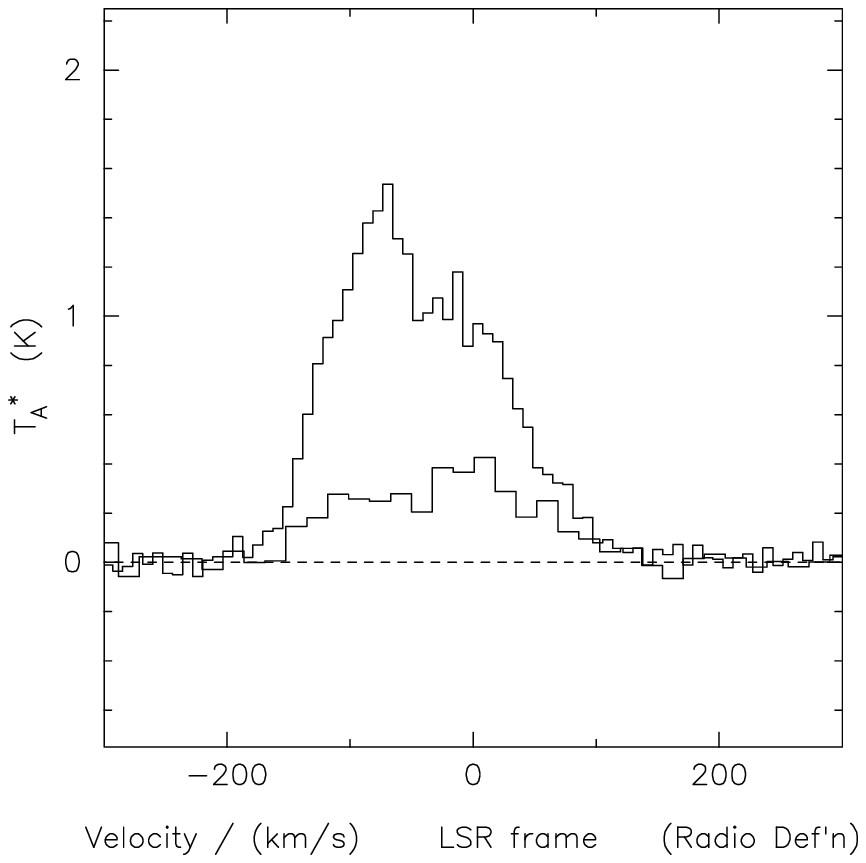}}
\end{minipage}
\hfill
\begin{minipage}[t]{3.94cm}
\resizebox{4.2cm}{!}{\includegraphics*{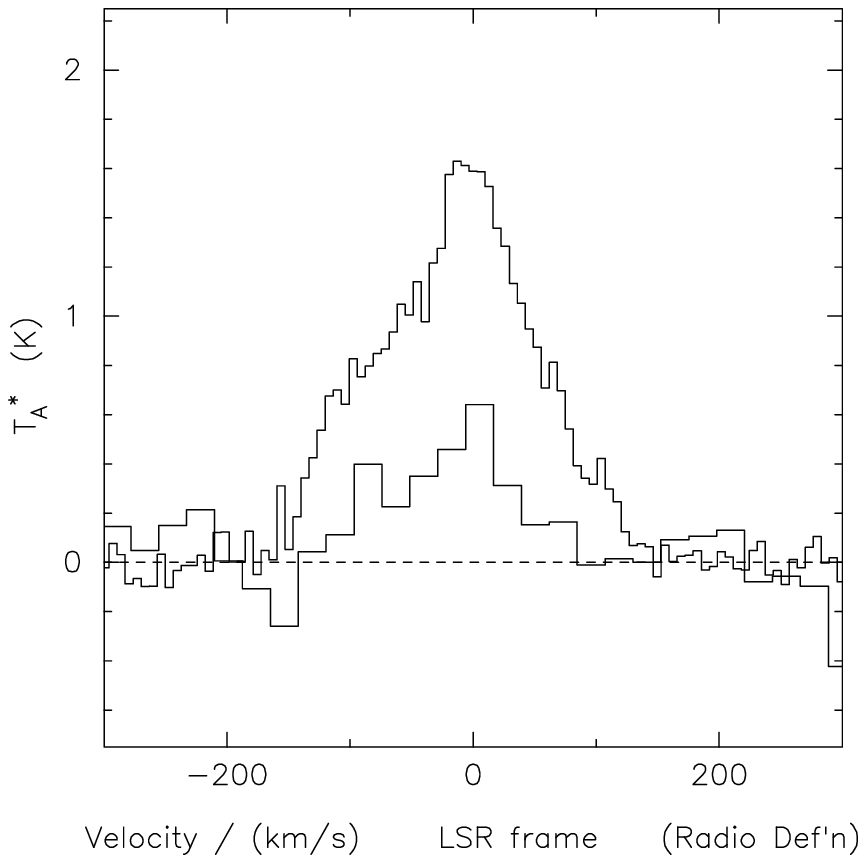}}
\end{minipage}
\hfill
\begin{minipage}[t]{3.94cm}
\resizebox{4.2cm}{!}{\includegraphics*{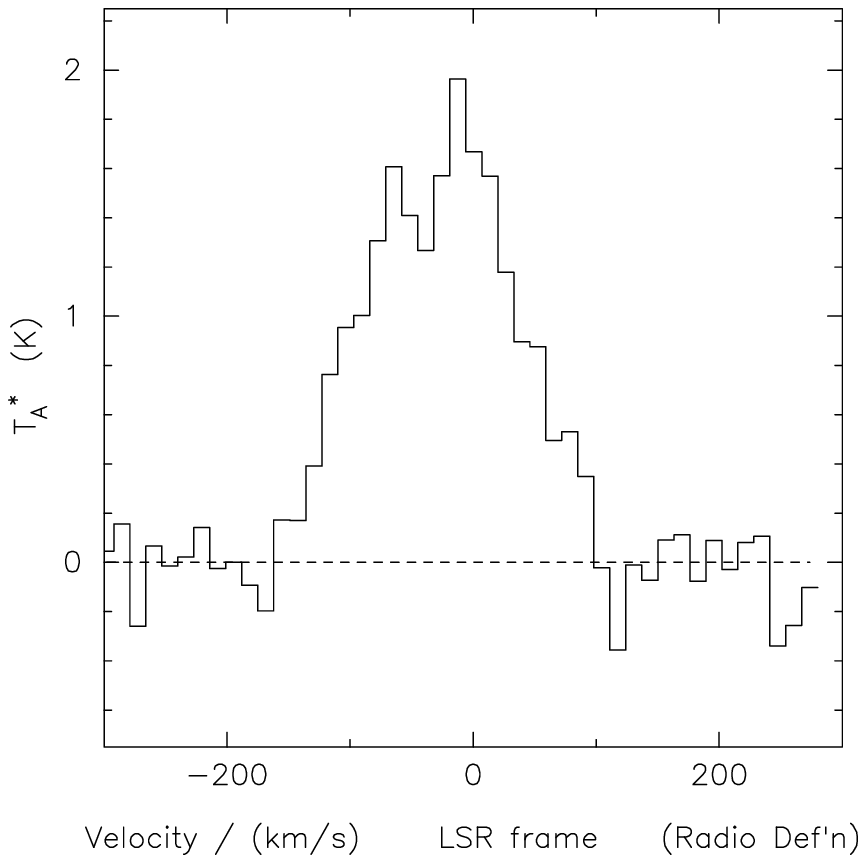}}
\end{minipage}
\hfill
\begin{minipage}[t]{3.94cm}
\resizebox{4.2cm}{!}{\includegraphics*{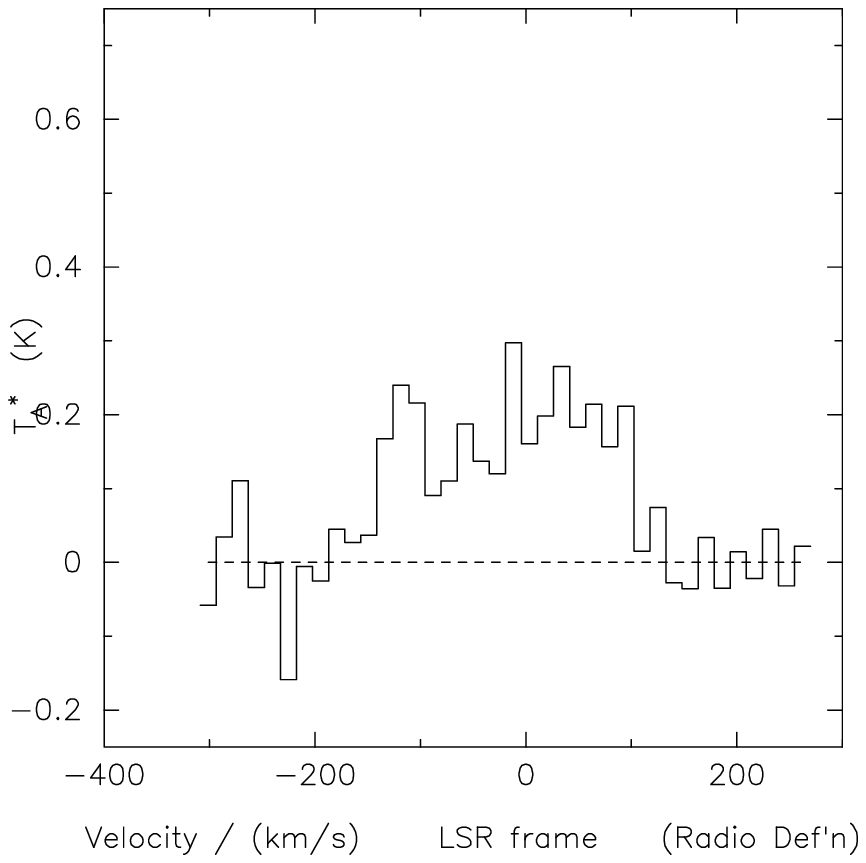}}
\end{minipage}
\caption[]
{Full resolution emission-line spectra observed towards the centers of 
IC~342 and Maffei~2.
Top row: IC~342; bottom row: Maffei~2. Columns from left to right:
$J$=2--1 CO, $J$=3--2 CO, $J$=4--3 CO, [CI]. Vertical scale is actually 
in $T_{\rm mb}$. Whenever available, $\13co$ profiles are shown as the 
lower of the two profiles in the appropriate $\co$ box. Their 
brightness temperatures have been multiplied by three, so that they
are depicted on the same temperature scale as [CI] in the right hand boxes.}
\end{figure*}

\section{Observations}

All observations described in this paper were carried out with the 15m 
James Clerk Maxwell Telescope (JCMT) on Mauna Kea (Hawaii) \footnote{The 
James Clerk Maxwell Telescope is operated on a joint basis between the 
United Kingdom Particle Physics and Astrophysics Council (PPARC), the 
Netherlands Organisation for Scientific Research (NWO) and the National 
Research Council of Canada (NRC).}. Details are given in Table 2.
Up to 1993, we used a 2048 channel AOS backend covering a band of 500 
MHz ($650\kms$ at 230 GHz). After that year, the DAS digital autocorrelator 
system was used in bands of 500 and 750 MHz. Integration times given in 
Table 2 are typical values used in mapping; central positions were usually
observed more than once and thus generally have significantly longer
integration times. Values listed are on+off. When sufficient free baseline was
available, we subtracted second or third order baselines from the profiles.
In all other cases, linear baseline corrections were applied. All spectra 
were scaled to a main-beam brightness temperature, $T_{\rm mb}$ = 
$T_{\rm A}^{*}$/$\eta _{\rm mb}$; relevant values for $\eta _{\rm mb}$ 
are given in Table 2. Spectra of the central positions in both galaxies are 
shown in Fig. 1 and summarized in Table 3.  In Table 2, we have also listed 
the parameters describing the various maps obtained. All maps are close to 
fully sampled. In all maps, the mapping grid was rotated by the angle 
given in Table 2 so that the Y axis coincided with the galaxy major axis. 
The velocity-integrated maps shown in Figs. 2 and 3 have been rotated back, so 
that north is (again) at top and the coordinates are right ascension and 
declination. For IC~342, the map grid origin in the $J$=2--1 and $J$=3--2
CO maps is within a few arcseconds from the optical centre listed in Table 1, 
whereas the $J$=4--3 and [CI] maps find their origin close to the radio 
centre; the two sets of maps are thus offset from one another by 
$\Delta \alpha$, $\Delta \delta$ = +9$''$, -3$''$. For Maffei~2, the grid 
origin is again close to the optical centre; the radio centre occurs in the
maps nominally at $\Delta \alpha$, $\Delta \delta$ = -0.8$''$, +6$''$.

%Figure2: Contour Maps of IC~342
\begin{figure*}
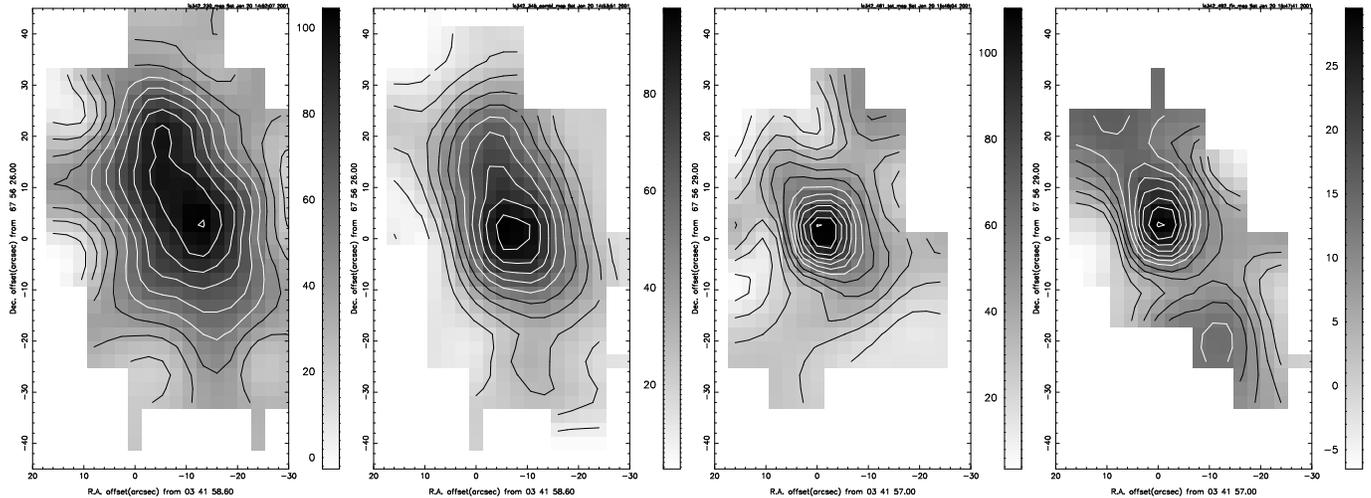

\unitlength1cm
\begin{minipage}[b]{4.4cm}
\resizebox{4.4cm}{!}{\includegraphics*{Fig2a.ps}}
\end{minipage}
\hfill
\begin{minipage}[t]{4.4cm}
\resizebox{4.4cm}{!}{\includegraphics*{Fig2b.ps}}
\end{minipage}
\hfill
\begin{minipage}[b]{4.4cm}
\resizebox{4.4cm}{!}{\includegraphics*{Fig2c.ps}}
\end{minipage}
\hfill
\begin{minipage}[b]{4.4cm}
\resizebox{4.4cm}{!}{\includegraphics*{Fig2d.ps}}
\end{minipage}
\caption[]
{Contour maps of emission from IC~342 integrated over the velocity
range $V_{\rm LSR}$ = -100 to 200 $\kms$. North is at top. Offsets
are marked in arcminutes with respect to J2000 coordinates 
$\alpha_{\circ} = 03^{h}46^{m}49.^{s}7, \delta_{\circ} =
+68^{\circ}05^{'}45^{''}$. Left to right: CO $J$=2--1, CO $J$=3--2, 
CO $J$=4--3 and [CI]. Contour values are linear in $\int T_{\rm mb} 
{\rm d}V$. Contour steps are 16 $\kkms$ (2--1, 3--2, 4--3) and 5 
$\kkms$ (CI) and start at step 1. }
\end{figure*}

%Figure3: Contour Maps of Maffei~2
\begin{figure*}
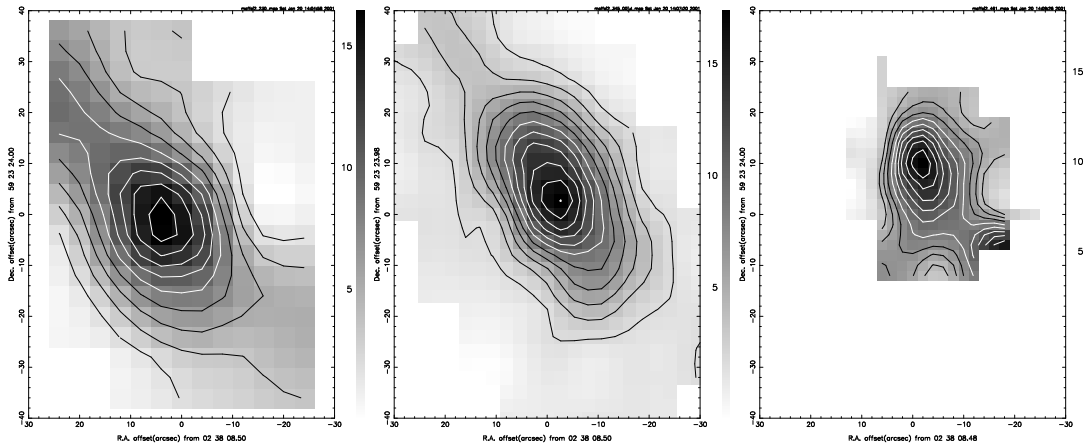

\begin{center}
\unitlength1cm
\begin{minipage}[b]{4.75cm}
\resizebox{4.75cm}{!}{\includegraphics*{Fig3a.ps}}
\end{minipage}
\begin{minipage}[t]{4.75cm}
\resizebox{4.75cm}{!}{\includegraphics*{Fig3b.ps}}
\end{minipage}
\begin{minipage}[b]{4.75cm}
\resizebox{4.75cm}{!}{\includegraphics*{Fig3c.ps}}
\end{minipage}
\caption[]
{Contour maps of emission from Maffei~2 integrated over the velocity range 
$V_{\rm LSR}$ = -200 to +150 $\kms$. North is at top.  Offsets are marked 
in arcminutes with respect to J2000 coordinates $\alpha_{\circ} = 
02^{h}41^{m}54.^{s}6, \delta_{\circ} = +59^{\circ}36^{'}11^{''}$. 
Left to right: CO $J$=2--1, CO $J$=3--2, CO $J$=4--3. Contour values are 
linear in $\int T_{\rm mb} {\rm d}V$. Contour steps are 25 $\kkms$ (2--1 and
3--2) and 35 $\kkms$ (4--3) and start at step 1. }
\end{center}
\end{figure*}

%Table 3
\begin{table*}[t]
\caption[]{Central CO and CI line intensities in IC~342 and Maffei~2}
\begin{flushleft}
\begin{tabular}{llrrccrcc}
\hline
\noalign{\smallskip}
& & & \multicolumn{3}{l}{IC~342} & \multicolumn{3}{l}{Maffei~2} \\
\multicolumn{2}{l}{Transition} & Resolution  
& $T_{\rm mb}$$^a$ & $\int T_{\rm mb}$d$V^a$ & $L_{\rm tot}$$^b$ 
& $T_{\rm mb}$$^a$ & $\int T_{\rm mb}$d$V^a$ & $L_{\rm tot}$$^b$ \\
&       & ($\arcsec$)       & (mK) & ($\kkms$)  & $\kkms$ kpc$^{2}$
			    & (mK) & ($\kkms$)  & $\kkms$ kpc$^{2}$ \\
\noalign{\smallskip}
\hline
\noalign{\smallskip}
$\co$   & $J$=2--1 & 21 & 3090 & 172$\pm$19 & 15.4 & 1500   & 245$\pm$35 & 11.6 \\
        &	   & 43 &      &  98$\pm$14 &	   &	    & 106$\pm$15 &  \\
        & $J$=3--2 & 14 & 3140 & 186$\pm$23 & 10.4 & 1650   & 295$\pm$35 &  7.6 \\
        &	   & 21 &      & 126$\pm$13 &	   &	    & 165$\pm$15 &  \\
        & $J$=4--3 & 11 & 4020 & 209$\pm$21 &  4.6 & 1950   & 405$\pm$50 &  5.7 \\
        &	   & 14 &      & 176$\pm$19 &	   &	    & 280$\pm$35 &  \\
        &	   & 21 &      & 115$\pm$14 &	   &	    & 160$\pm$25 &  \\
\noalign{\smallskip}
$\13co$ & $J$=2--1 & 21 &  476 & 24.0$\pm$3 & ---  &  150   & 22$\pm$4  &   --- \\
	&	   & 43 &      & 12.0$\pm$2 &	   &	    &	---     &   \\
	& $J$=3--2 & 14 &  311 & 17.1$\pm$2 & ---  &  200   & 20$\pm$4  &   ---\\
	&	   & 21 &      & 14.4$\pm$2 &      &        &	---     &   \\
\noalign{\smallskip}
[CI] & $^{3}$P$_{1}$--$^{3}$P$_{0}$ 
                   & 10 & 1030 & 54$\pm$6   &  1.0 &  190   & 37$\pm$7  & --- \\
	&          & 14 &      & 42$\pm$8   &      &        &   ---     & \\
	&          & 21 &      & 27$\pm$7   &      &        &   ---     & \\
\noalign{\smallskip}
\hline
\end{tabular}
\end{flushleft}
Notes to Table 3: a. Beam centered on nucleus; b. Total central concentration
\end{table*}

%Table 4  Line Ratios
\begin{table*}
\begin{center}
\caption[]{\centerline{Integrated line ratios in the centres of IC~342 and Maffei~2}}
\begin{tabular}{lcccccc}
\hline
\noalign{\smallskip}
Transitions & \multicolumn{2}{l}{IC~342} & \multicolumn{2}{l}{Maffei~2} \\
	    & Nucleus & Total Center     &  Nucleus & Total Center  \\ 
\noalign{\smallskip}
\hline
\noalign{\smallskip}
$\co$ (1--0)/(2--1)$^{a}$    & 0.95$\pm$0.1  & 0.9  & 0.9$\pm$0.2    & 0.8: \\
$\co$ (3--2)/(2--1)$^{b}$    & 0.73$\pm$0.10 & 0.7  & 0.75$\pm$0.13  & 0.7 \\
$\co$ (4--3)/(2--1)$^{b}$    & 0.67$\pm$0.10 & 0.3  & 0.65$\pm$0.10  & 0.5 \\
\noalign{\smallskip}
$\co$/$\13co$ (1--0)$^{c}$ & 10.7$\pm$1.3    & ---  &  8.6$\pm$1.1   & ---  \\
$\co$/$\13co$ (2--1)$^{b}$ &  7.2$\pm$1.5    & ---  & 10.0$\pm$1.4   & ---  \\
$\co$/$\13co$ (3--2)$^{d}$ & 10.2$\pm$1.4    & ---  & 12.4$\pm$1.7   & ---  \\
\noalign{\smallskip}
CI/CO(2--1)$^{b}$   	   & 0.16$\pm$0.03   & 0.06 & 0.10$\pm$0.05  & --- \\
CII/CO(2--1)$^{e}$  	   & 0.54	     &	    & 0.31	     & \\
\noalign{\smallskip}
\hline
\end{tabular}
\end{center}
Notes: 
a. IC~342 ratio from Eckart et al. (1990), Xie et al. (1994) and Meier
   et al. (2000); Maffei~2 ratio from $J$=1--0 data by Weliachew et al. 
   (1988); both ratios at 21$''$ resolution. 
b. This Paper, JCMT at 21$''$ resolution;
c. Weighted mean based on ratios presented by Rickard $\&$ Blitz (1985); 
   NRAO at 65$''$ resolution; Young $\&$ Sanders (1986); FCRAO at 
   45$''$ resolution; Weliachew et al. (1988); IRAM 30 m at 24$''$ 
   resolution; Sage $\&$ Isbell (1991); NRAO 12m at 57$''$ resolution; 
   Xie et al. (1994) FCRAO at 45$''$ resolution; Meier et al. (2000) OVRO at 
   4.$''$5 resolution
d. This Paper; JCMT at 14$''$ resolution.
e. From Crawford et al. (1985) and Stacey et al. (1991), KAO at 55$''$ 
   resolution; JCMT convolved to 55$''$ resolution.
\end{table*}

%Figure4: Major Axis pv maps IC~342
\begin{figure*}
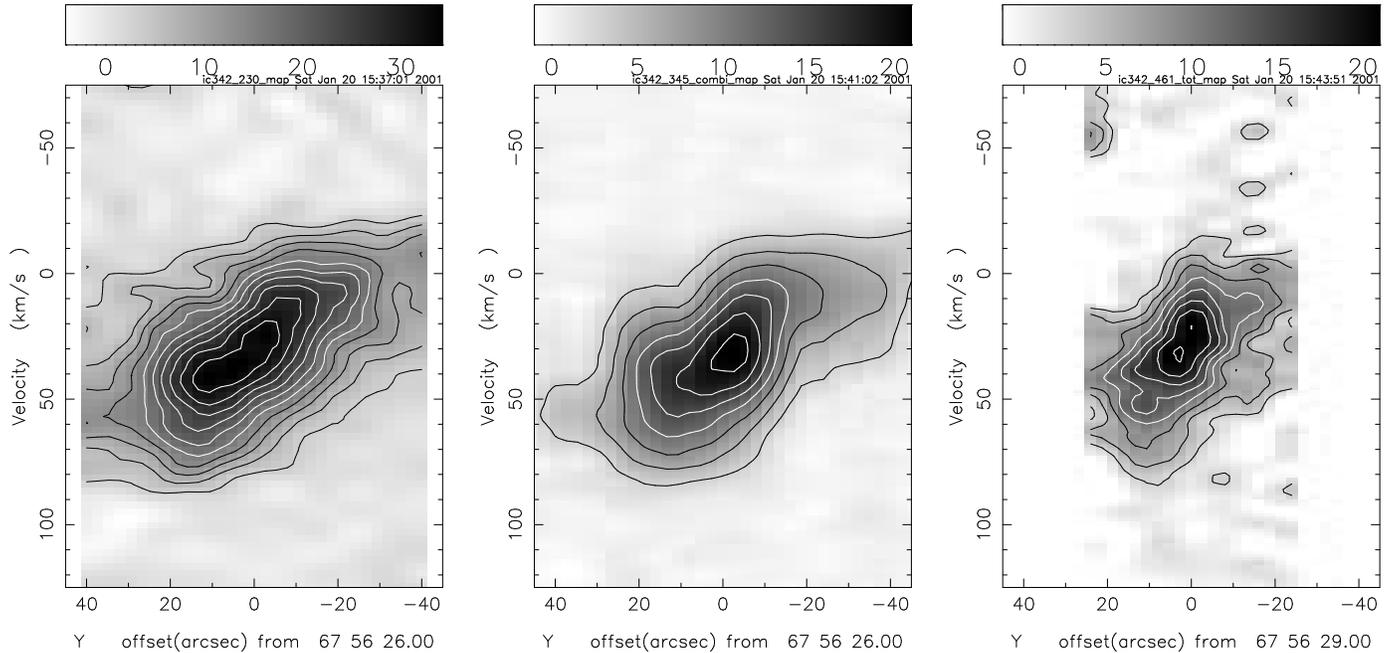

\unitlength1cm
\begin{minipage}[b]{5.54cm}
\resizebox{5.8cm}{!}{\includegraphics*{Fig4a.ps}}
\end{minipage}
\hfill
\begin{minipage}[t]{5.54cm}
\resizebox{5.8cm}{!}{\includegraphics*{Fig4b.ps}}
\end{minipage}
\hfill
\begin{minipage}[t]{5.54cm}
\resizebox{5.8cm}{!}{\includegraphics*{Fig4c.ps}}
\end{minipage}
\caption[]
{Position-velocity maps of CO emission from IC~342 in position angle
20$^{\circ}$. Left to right: CO $J$=2--1, CO $J$=3--2 and CO $J$=4--3. 
Contour values are linear in $T_{\rm mb}$. All maps are one beamwidth
wide, have a velocity resolution of 7.5 km s$^{-1}$ and have contour 
steps of 5 K, starting at step 1. Vertical scale is $V_{\rm LSR}$
}
\end{figure*}

%Figure 5: Major axis pv maps M~83
\begin{figure*}
\unitlength1cm
\begin{minipage}[b]{5.54cm}
\resizebox{6cm}{!}{\includegraphics*{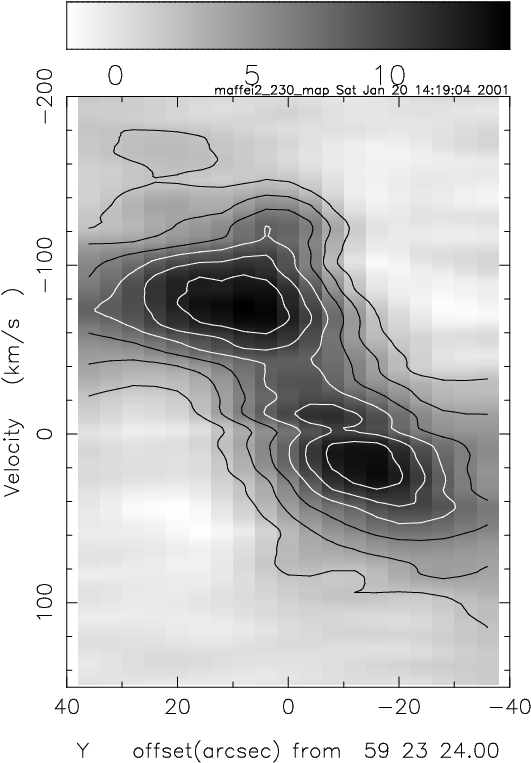}}
\end{minipage}
\hfill
\begin{minipage}[t]{5.54cm}
\resizebox{6cm}{!}{\includegraphics*{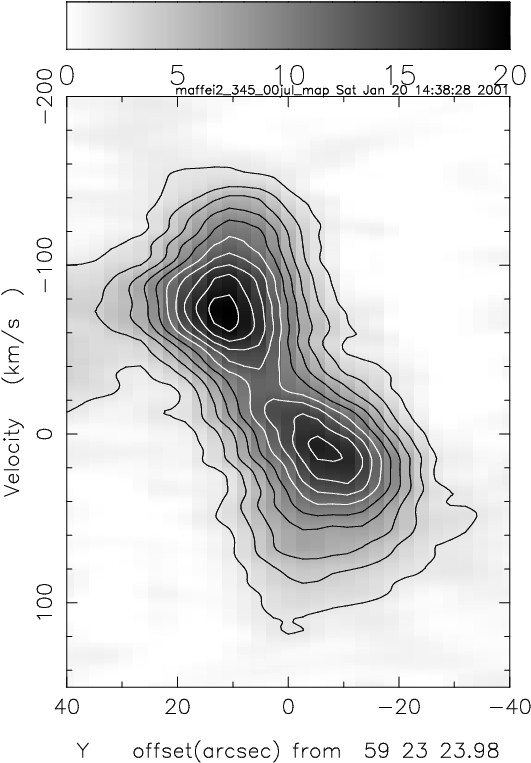}}
\end{minipage}
\hfill
\begin{minipage}[t]{5.54cm}
\resizebox{6cm}{!}{\includegraphics*{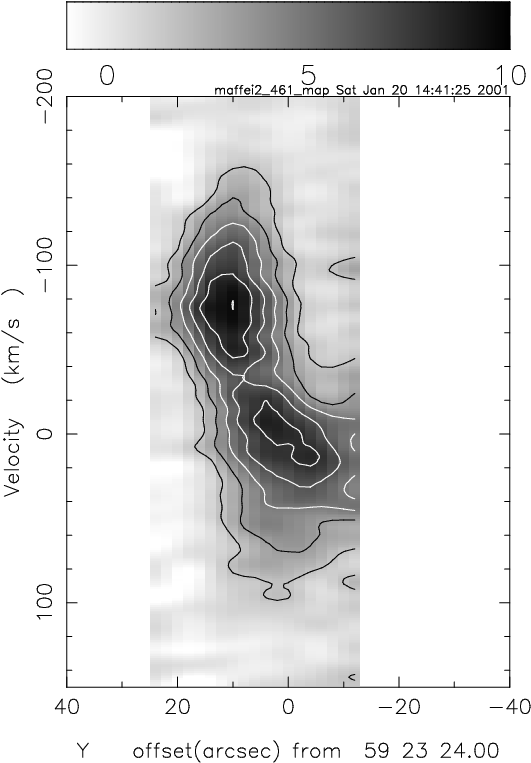}}
\end{minipage}
\caption[]
{Position-velocity maps of CO emission from Maffei~2 in position angle 
25$^{\circ}$. Left to right: CO $J$=2--1, CO $J$=3--2 and CO $J$=4--3. 
Contour values are linear in $T_{\rm mb}$. All maps are one beamwidth
wide, have a velocity resolution of 10 km s$^{-1}$ and have contour 
steps of 3 K, starting at step 1. Vertical scale is $V_{\rm LSR}$
}
\end{figure*}

\section{Results}

\subsection{CO distribution}

In both galaxies, molecular gas as traced by CO is strongly
concentrated towards the center. The central molecular source is
contained with $R <$ 250 pc in the case of IC~342, and $R < 200
\times D/2.7$ pc in the case of Maffei~2, where $D$ is the true
distance of Maffei~2 in Mpc. 

CO line aperture synthesis maps of IC~342 show, at resolutions of 
2$''$--4$''$, considerable structure only hinted at in our maps 
(Fig.~2). Clearly, the central CO concentration is not toroidal
in shape. Rather, the maps published by Sakamoto et al. (1999) and 
Meier et al. (2000) show individual CO peaks (cloud complexes) 
distributed along a perspectively foreshortened double spiral --
although the foreshortening seems more than expected for an
inclination of only 25$^{\circ}$. 
The molecular gas enhancement of spiral arms ends at about $R$ = 
250 pc from the nucleus. The nucleus itself is located at a minimum 
in the CO distribution. Emission from the high-density tracer HCN 
offers a very similar appearance (Downes et al. 1992). Our $J$=2--1 
$^{12}$CO map shows an elongated source, in which a smoothed version 
of the high-resolution map by Meier et al. (2000) is easily recognized. 
Emission in all our CO maps, as well as the [CI] map peaks at
the position coinciding with cloud complex B in the designation
by Downes et al. (1992). This peak becomes more pronounced with
increasing $J$ level, and in [CI]. This is partly the consequence of
increasing angular resolution, but we also note that the peak
coincides with the strongest concentration of thermal radio continuum 
emission (Turner $\&$ Ho 1983), thus with the region containing the 
highest spatial density of early type stars. The peak line intensities 
in Table~3 and the line ratios in Table~4 refer to this maximum. We 
thus expect them to exhibit properties commensurate with a starburst 
environment, i.e. those of a photon-dominated region (PDR).

The region corresponding to complexes C and D (Downes et al. 1992),
referred to as `Eastern Ridge' by Eckart et al. (1990), behaves
differently. It is almost as pronounced as region B in the $J$=2--1
transition, but fades rapidly at higher $J$ levels. Its molecular gas
must thus be cooler on the whole than that of region B. This is,
in any case, consistent with the much less impressive appearance
of this region in the radio continuum maps by Turner $\&$ Ho (1983).
There appear to be relatively weak, secondary maxima in the neutral
carbon emission at the northern and southern edges of the main
CO distribution, in addition to the primary [CI] maximum at cloud B.

The major-axis position-velocity map of IC~342 covers a relatively 
small velocity range, in accordance with the mostly face-on 
orientation of the galaxy. With increasing $J$ level, hence
increasing resolution and decreasing beamsmearing, the rotation 
steepens. In the $J$=4--3 CO p-V map the central material covers 
a range of about 80 km s$^{-1}$ over about 20$''$. The rotation in 
this region is characterized by a velocity gradient (cf. $J$=4--3 
CO in Fig. 4 and $J$=1--0 CO in Fig. 4 of Sakamoto et al. 1999) 
d$V$/d$\theta$ = 7.8 $\kms$/$''$ , corresponding to d$V$/d$R 
\approx$ 2.1 $\kms$/pc in the plane of the galaxy.

The structure of the central CO source in Maffei~2 is more complex
than Fig.~3 suggests. The galaxy is much more tilted than IC~342, 
and the integrated velocity images, showing only a resolved but 
featureless central source, are misleading. Maps of peak temperature 
$T_{\rm mb}$ rather than integrated-velocity temperature maps show
a distinct double-peaked source (see also Hurt et al. 1993). This is
in sharp contrast to IC~342, where peak-temperature and 
integrated-temperature maps show only marginally different morphologies.
The deceptively simple appearance of Maffei~2 in Fig.~3 is easily
explained by a glance at the major axis position-velocity maps
in Fig.~5. The single peak in the integrated map results from the
line-of-sight superposition of two distinct velocity components.
This is, in fact, the same situation that we found to apply to NGC~253
(Israel et al. 1995), and it is one to watch out for in all 
high-inclination galaxies. The CO source in Fig. 3 becomes more compact
with increasing $J$ level, i.e. increasing frequency. As can be seen in
Fig.~5, this is mostly the consequence of increasing resolution and
decreasing beamsmearing. In the peak-temperature maps (not shown) the
separation of the two peaks is only 13$''$ (corresponding to 170 pc).
The CO distribution shown in Figs.~3 and 5 is characteristic for a
toroidal distribution of molecular gas, with a 45 -- 60 pc wide
gap in the center. However, the `S'-shaped morphology of CO in
high-resolution (6$''$) aperture synthesis maps by Ishiguro et al.
(1989) and Hurt $\&$ Turner (1991) strongly suggests greatly
enhanced emission from molecular gas in inner spiral arms instead.

The effects of beamsmearing are quite noticeable in the p-V maps in 
Fig. 5. The average velocity gradient increases from 4.5 $\kms/''$ 
in the $J$=2--1 CO map to about 18 $\kms/''$ in the $J$=4--3 CO map.
The latter value corresponds to d$V$/d$R$ = 1.4 $\kms$/pc, i.e. not 
much different from the gradient in IC~342 (or NGC~6946 and M~83 for 
that matter -- see Israel $\&$ Baas 2001). We note that the CO
distribution in Fig.~5 just fills the steep leading edge of the
major axis HI distribution in Maffei~2 (Fig.~9 in Hurt et al. 1996).

\subsection{Line ratios}

We have determined the intensity ratio of the observed transitions at 
the centers of both galaxies; all intensities are normalized to 
those of the $J$=2--1 $\co$ line. The columns in Table 4 marked 
`Nucleus' refer to ratios in a 21$''$ beam centered on the nucleus,
with the exception of the [CII]/CO(2--1) ratio which was determined 
after convolution of the $J$=2--1 $^{12}$CO measurements to the
55$''$ beam of the [CII] observations, and the $J=1-0$ $^{12}$CO/$^{13}$CO
isotopic ratio which represents a weighted mean from a number of
determinations at various resolutions. The [CII] intensities (Crawford 
et al. 1985; Stacey et al. 1991), were converted to velocity-integrated 
temperatures to obtain the line ratios in Table 4. The columns marked `Total 
Center' refer to the intensities of the central concentration 
integrated over the source extent in the maps. The apparent decrease in
source extent going from the lower $J$ levels (lower frequencies) to
the higher frequencies is mostly an artifact of limited resolution.
When corrected for finite beamwidth, source dimensions at e.g. the
$J$=2--1 and $J$=4--3 transitions are in fact very similar. Nevertheless, 
the smaller area coverage in the [CI] and $J$=4--3 CO maps may ahve led 
to an underestimate of the total intensity. The entries in Table 4 
suggest that this may indeed be the case.
The $J$=1--0/$J$=2--1 ratios have relatively large uncertainties,
because we could not use observations of our own, but instead
have used estimated $J$=1--0 intensities reduced to a 21$''$ beam 
from the references given in the table. These ratios are, in any case, 
close to unity.

The $^{12}$CO ratios are quite similar for both galaxies, but the 
intensity of the [CI] and [CII] lines relative to CO is significantly 
lower in Maffei~2 than in IC~342, suggesting stronger PDR effects in 
the latter. The intensity of the [CII] line indicates the presence
of both high temperatures and high gas densities in the medium as the
critical values for this transition are $T_{\rm kin} \geq$ 91 K and
$n \geq 3500 \cc$. At the same time, such values must be reconciled
with the much lower temperatures and (column) densities implied by
the more modest isotopic ratios in the lower CO transitions. These 
isotopic ratios differ for the two galaxies. For each individual
transition the isotopic ratios just agree within the errors, but the
overall patterns are clearly different. For Maffei~2, the isotopic ratio
increases with increasing $J$ level, but for IC~342 the isotopic
ratio is minimal in the $J$=2--1 transition. These differences
suggest significant differences in the molecular gas properties
at the positions sampled in both galaxies

\section{Analysis}

\subsection{Modelling of CO}

%Table 5
\begin{table*}
\caption[]{Model parameters}
\begin{flushleft}
\begin{tabular}{lccccccc}
\hline
\noalign{\smallskip} 
Model & \multicolumn{3}{c}{Component 1}    	  & \multicolumn{3}{c}{Component 2}  		  & Relative\\
     & Kinetic	    & Gas	 & CO Column   	  & Kinetic 	& Gas 	      	   & CO Column    & $J=2-1 ^{12}$CO Emission \\
     & Temperature  & Density    & Density     	  & Temperature & Density     	   & Density      & Component 1:2 \\
     & $T_{\rm k}$  & $n(\h2$) & $N$(CO)/d$V$	  & $T_{\rm k}$ & $n(\h2$)  	   & $N$(CO)/d$V$ & \\
     & (K)     	    & ($\cc$)    & ($\cm2 (\kms)^{-1}$) & (K) 	& ($\cc$ & ($\cm2 (\kms)^{-1}$)  & \\
\noalign{\smallskip}
\hline
\noalign{\smallskip}
\multicolumn{8}{c}{IC~342}\\
\noalign{\smallskip}
\hline
\noalign{\smallskip}
1 &   150    &  1000      &      3$\times10^{17}$ &   150       &   3000     &        0.6$\times10^{17}$ & 0.65 : 0.35 \\
2 &   100    &  1000 	  &      3$\times10^{17}$ &   100    	&   3000     &          1$\times10^{17}$ & 0.35 : 0.65 \\
3 &    10    & 10000	  &      3$\times10^{17}$ &   100    	&   3000     &          1$\times10^{17}$ & 0.25 : 0.75 \\
\noalign{\smallskip}
\hline
\noalign{\smallskip}
\multicolumn{8}{c}{Maffei~2} \\
\noalign{\smallskip}
\hline
\noalign{\smallskip}
4 &   150    &   500      &     10$\times10^{17}$ &   150       &   3000     &        0.3$\times10^{17}$ &  0.25 : 0.75 \\
5 &   150    &   100      &     10$\times10^{17}$ &    60       &  10000     &        0.6$\times10^{17}$ &  0.50 : 0.50 \\
6 &   100    &   100 	  &     10$\times10^{17}$ &    20       & 100000     &        0.3$\times10^{17}$ &  0.30 : 0.70 \\
7 &    10    &  3000 	  &      6$\times10^{17}$ &   150       &   3000     &        0.6$\times10^{17}$ &  0.35 : 0.65 \\
\noalign{\smallskip}
\hline
\end{tabular}
\end{flushleft}
\end{table*}

%Table 6
\begin{table*}
\caption[]{Beam-averaged results}
\begin{center}
\begin{tabular}{lcccccccc}
\hline
\noalign{\smallskip} 
Model & \multicolumn{3}{c}{Beam-Averaged Column Densities} & \multicolumn{2}{c}{Total Central Mass} & \multicolumn{2}{c}{Face-on Mass Density}       & Relative Mass \\
   & $N$(CO) & $N$(C)    & $N(\h2)$ & $M(\h2)$  & $M_{\rm gas}$        & $\sigma(\h2)$ & $\sigma_{\rm gas}$ & Components 1:2 \\
   & \multicolumn{2}{c}{($10^{18} \cm2$)} & ($10^{21} \cm2$) & \multicolumn{2}{c}{($10^{7} \Msun$)} & \multicolumn{2}{c}{($\Msun$/pc$^{-2}$)} & \\
\noalign{\smallskip}
\hline
\noalign{\smallskip}
\multicolumn{9}{c}{IC~342; $N_{\rm H}/N_{\rm C}$ = 2200; $N$(HI)$^{a} = 1.5\times 10^{20} \cm2$} \\
\noalign{\smallskip}
\hline
\noalign{\smallskip}
1  & 0.8 & 2.3      & 3.3      &  0.5 &  0.6 &  49 &  67 &  0.7 : 0.3 \\
2  & 0.7 & 3.8      & 4.8      &  0.7 &  0.9 &  70 &  96 &  0.4 : 0.6 \\
3  & 0.7 & 1.0      & 4.9      &  0.7 &  1.0 &  72 &  99 &  0.3 : 0.7 \\
\noalign{\smallskip}
\hline
\noalign{\smallskip}
\multicolumn{9}{c}{Maffei~2; $N_{\rm H}/N_{\rm C}$ = 2500; $N$(HI)$^{b} = 1.1 \times 10^{21} \cm2$} \\
\noalign{\smallskip}
\hline
\noalign{\smallskip}
4  & 1.7 & 1.6      & 3.5      &  0.6 &  0.9 &  28 &  43 &  0.55 : 0.45 \\
5  & 3.6 & 1.3      & 5.5      &  0.9 &  1.4 &  43 &  64 &  0.8 : 0.2 \\
6  & 2.6 & 1.1      & 4.1      &  0.7 &  1.1 &  32 &  49 &  0.7 : 0.3 \\
7  & 2.0 & 1.9      & 4.4      &  0.7 &  1.1 &  32 &  53 &  0.55 : 0.45 \\
\noalign{\smallskip}
\hline
\end{tabular}
\end{center}
Notes: a. Newton (1980); b. Hurt et al. (1996)
\end{table*}

The observed $\co$ and $\13co$ line intensities and ratios have been 
modelled with the large-velocity gradient (LVG) radiative transfer 
models described by Jansen (1995) and Jansen et al. (1994). These 
provide model line intensities as a 
function of three input parameters: gas kinetic temperature $T_{\rm k}$, 
molecular hydrogen density $n(H_{2})$ and CO column density per unit 
velocity ($N({\rm CO})$/d$V$). By comparing model line {\it ratios} to the 
observed ratios we have identified the physical parameters best describing 
the actual conditions in the observed source. Beam-averaged properties
are determined by comparing observed and model intensities. In principle,
with seven measured line intensities, properties of a single gas component
are overdetermined as only five independent observables are required.
We found that no single-component fit could be made to the data of 
either IC~342 or Maffei~2.

However, good fits based on two gas components can be obtained. In 
order to reduce the number of free parameters, we assume identical CO 
isotopical abundances for both gas components and assign the specific 
value [$^{12}$CO]/[$^{13}$CO] = 40 (Mauersberger $\&$ Henkel 1993;
Henkel et al. 1998). 
We identified acceptable fits by searching a grid of model parameter 
combinations (10 K $\leq T_{\rm k} \leq $ 250 K, $10^{2} \cc \leq
n(\h2) \leq 10^{5} \cc$, $6 \times 10^{15} \cm2 \leq N(CO)/dV \leq 
3 \times 10^{18} \cm2$) for model line ratios matching the observed set,
with the relative contribution of the two components as the single
free parameter. Largely as a consequence of the non-negligible finite
errors in the line ratios, solutions are not unique, but rather delineate
a range of values in a particular region of parameter space. To a
certain extent, variations in input parameters may compensate for
one another, leading to identical line ratios for somewhat different
combinations of input parameters. We have rejected all solutions in
which the denser gas component is also hotter than the more tenuous
component, as we consider this physically unlikely on the linear scales
observed. From the remainder of solutions, we have selected characteristic
examples and listed these in Table~5.

The line ratios observed for IC~342 allow only a rather limited range of
gas parameters. One of the two components has well-determined parameters:
a column density $N({\rm CO})$/d$V$ = 6--10 $\times 10^{16} \cm2$, an $\h2$ 
density of 3000 $\cc$) and a kinetic temperature of 100 -- 150 K. 
The other gas component either is less dense and equally hot, or denser 
and much cooler with a temperature of the order of 10 -- 20 K.
For Maffei~2, the choice of models is less restricted. Possible
solutions pair a low-density, hot component and rather large column
densities with a much denser, low-column density component of uncertain
temperature. Alternatively, both components have a reasonably
high density of 3000 $\cc$, but represent either a low temperature/high 
column-density or a high temperature/low column-density gas.

A further check on the validity of the model results is provided by
the C$^{18}$O measurements obtained by Eckart et al. (1990). This
isotope was not included in our modelling requirements. For isotopic
ratios between $^{12}$CO/C$^{18}$O = 250 (model 1) and $^{12}$CO/C$^{18}$O 
= 225 (model 3), the parameters of our solutions reproduce the observed 
intensity ratios in the $J$=1--0 and $J$=2--1 transitions within 10$\%$.

\subsection{Beam-averaged molecular gas properties}

The relation between carbon monoxide, neutral and ionized carbon combines 
the observed C and C$^{+}$ intensities with the chemical models by Van 
Dishoeck $\&$ Black (1988) which show a strong dependence of the 
$N({\rm C})/N({\rm CO})$ column density ratio on total carbon and molecular 
hydrogen column densities. In the analysis of [CI], we took the kinetic 
temperatures, $\h2$ densities and filling factors resulting from the
CO analysis, and then solved for column density $N({\rm CI})$. In practice, 
the column density of one component is frequently well-determined, 
whereas that of the other is more or less degenerate. For this reason, 
we solved for identical velocity dispersions in the two gas components. 
This procedure we also followed for [CII]; in those cases where both
model kinetic temperature and gas density are very different for the
two components, we have modelled a {\it single} [CII] component with
the higher kinetic temperature and the higher gas density of the two
components. This applies specifically to models 5 and 6 (see below).
Finally, we have related total carbon (i.e. C + CO) column densities 
to molecular hydrogen column densities by using an estimated [C]/[H] 
gas-phase abundance ratio. 

IC~342 has a measured central abundance [O]/[H] = 2.0 $\times$ 10$^{-3}$ 
(Vila-Costas $\&$ Edmunds 1992, Garnett et al. 1997). We are unaware of
any abundance determination for Maffei~2, and therefore use the mean
value for the spiral galaxies listed by Zaritzky et al. (1994), 
[O]/[H] = 1.7 $\times$ 10$^{-3}$. Then, using the results obtained 
by Garnett et al. (1999), notably their Figs. 4 and 6, we have estimated
carbon abundances [C]/[H] = 1.7$\pm$0.5 $\times$ 10$^{-3}$ and 
1.45$\pm$0.5 $\times$ 10$^{-3}$ for IC~342 and Maffei~2 respectively.
As a significant fraction of carbon is tied up in dust particles and 
thus unavailable in the gas-phase, we have adopted a fractional 
correction factor $\delta_{\rm c}$ = 0.27 (see for instance van Dishoeck 
$\&$ Black 1988), so that $N_{\rm H}$ = [2$N(\h2) + N$(HI)] = {\it A}\, 
[$N$(CO) + $N$(C)], where {\it A} = 2200 and 2500 for IC~342 and Maffei~2
respectively. The numerical parameter {\it A} is uncertain by
about a factor of two. 

In Table~6, we present beam-averaged column densities for both CO and 
C (C$^{\rm o}$ and C$^{+}$), and $\h2$ column densities derived under
the assumptions just discussed. We also present the total masses 
estimated to be present in the central molecular concentration 
calculated from $L_{tot}$ listed in Table 3, and the face-on mass 
densities implied by hydrogen column density and the galaxy inclination. 
Notwithstanding the differences between the model cloud parameters, 
the beam-averaged results in Table~6 are rather similar: hydrogen 
column densities, masses and mass-densities are well within a factor 
of two from one another. Beam-averaged {\it neutral} carbon to carbon 
monoxide column density ratios range from $N$(C$^{\rm o}$)/$N$(CO) = 0.5--0.6 
for IC~342 to $N$(C$^{\rm o}$)/$N$(CO) = 0.15--0.3 for Maffei~2. These 
values are well within the range found for other late-type galaxies
such ad M~82, NGC~253, M~83, and NGC~6946 (White et al. 1994; Israel 
et al. 1995; Stutzki et al. 1997; Petitpas $\&$ Wilson 1998; Israel 
$\&$ Baas, 2001).

\subsection{The center of IC~342}

Even before the model fitting described in the previous section,
the center of IC~342 was identified as the abode of hot and dense
molecular gas, first by Martin et al. (1982) and Martin $\&$ Ho
(1985) who found the presence of gas with kinetic temperatures 
estimated to be 70 K or higher. Somewhat later, measurements
of the high-density tracer HCN led Downes et al. (1992) to suspect
the presence of molecular gas at various densities in the range 
$n(\h2) = 150 - 3 \times 10^{4} \cc$ and kinetic temperatures 
in the range $T_{\rm kin}$ = 50 -- 70 K. This gas consisted of
a few dense cloud complexes and much less dense gas inbetween.
Multitransition HCN measurements (Jackson et al. 1995; Paglione
et al. 1997) and other molecules (Paglione et al. 1995; 
Mauersberger et al 1995; H\"uttemeister et al 1997; Henkel et al
2000) suggest kinetic temperatures in the range of 50 - 200
K, densities of about 10$^{4} \cc$ as well as the presence of 
additional gas components at lower densities. Such temperatures and
densities are not unexpected for the site of a significant burst
of star formation (cf. Turner $\&$ Ho, 1983; 1994). The PDR model 
calculations by Kaufman et al. (1999), applied to the relative 
intensities of CO, [CI], [CII] and the far-infrared continuum of the
center of IC~342 likewise suggest a temperature $T_{\rm kin} 
\approx$ 150 K, a density around 10$^{4} \cc$ and an ambient 
radiation field strength log $G_{\rm o}$ = 2.5 ($G_{\rm o}$ 
expressed in units of the Habing Field, i.e. $1.6\times10^{-3}$ 
erg s$^{-1}$ cm$^{-2}$. All three 
model solutions in Table~5 are consistent with these values, 
although the higher densities suggested by the other molecular 
line observations are only provided by model 3. More support for 
model 3 follows from the C$^{18}$O aperture synthesis observations 
carried out by Meier $\&$ Turner (2001) who find that the molecular 
cloud complexes have densities of the order of 1000--3000 $\cc$ 
and, more importantly, kinetic temperatures of 10--40 K (see also 
Meier et al. 2000). We thus accept model 3 as the most probable 
approximation of the physical conditions in the central molecular 
gas of IC~342. About a third of the total 
molecular mass resides in a dense, fairly cold component with an
excitation temperature close to the kinetic temperature, and very
high optical depths in all observed CO transitions; even the $^{13}$CO 
transitions are marginally optically thick with $\tau$ = 1--2.
Most of the remaining mass is hot, with T$_{\rm ex}$ = 50 K. The $J$=1--0
$^{12}$CO transition is optically thin, but the higher transitions
are optically thick although optical depths are not extremely high.

Obviously, actual conditions will be more varied and complex, but even 
then we expect the beam-averaged results to be not very different from 
the ones obtained here. We find a total central gas mass of about 
10$^{7}$ $\Msun$ which agrees well with estimates derived
from e.g. HCN observations, but is much lower than the mass values
quoted in previous CO studies. However, these were all based on
{\it assumed} CO to $\h2$ ratios, notably the $\h 2$ column density
to CO intensity ratio $X$. The $\h2$ column densities and masses
presented in this paper are, in contrast, based on total gas-phase 
carbon amounts and the actual carbon abundance. The uncertainty in
these results is essentially determined by the error in the
product $\delta_{C}\, \times$ [C]/[H], which is hard to quantify, but 
which we estimate to be about a factor of two. A mass of $10^{7}$ 
$\Msun$ is not very high, yet it is about 10$\%$ of the dynamical
mass with $R$ = 130 pc; this renders any value much higher rather
unlikely. Model 3 implies for IC~342 a value $X = 3 \times 10^{19}$
$\h2$ mol cm$^{-2}$/$\kkms$, i.e. almost an order of magnitude lower
than commonly used $X$ values for the Galactic Solar Neighbourhood.
However, such a low value is not uncommon for a galactic center.
We have found similar values for NGC~6946 and M~83 (Israel $\&$ Baas 
2001), and such values are by now solidly established for the center 
of our own Milky Way (Sodroski et al. 1995; Oka et al 1998; Dahmen
et al. 1998). As an important consequence of these low values for $X$ 
and beam-averaged $N(\h2)$, the {\it abundances of all molecular species} 
that were derived using erroneously estimated much higher $\h2$ column
densities {\it must be revised upwards} by almost an order of magnitude.
This is particularly important in the case of IC~342. Because it is 
one of the few galaxies bright in molecular line emission, it has 
extensively been used as the basis for abundance determinations of
a relatively great variety of molecular species. 

\subsection{The center of Maffei~2}

In a similar vein, rough estimates of conditions to be found in
the center of Maffei~2 follow from observations of HCN, CS and
other molecules: $\h2$ densities in the range $10^{4} - 10^{5} \cc$
and probably closer to the former than to the latter, as well as kinetic
temperatures of about 100 K (Mauersberger et al. 1989; H\"uttemeister
et al. 1997; Paglione et al. 1997). The relative intensities
of CO, CI, CII and the far-infrared continuum imply, within the
context of the PDR model by Kaufman et al. (1999), somewhat higher
densities close to 10$^{5} \cc$, temperatures of about 150 K and a
radiation field log $G_{\rm o}$ = 2.3. These temperatures and 
densities suggest that models 5 and 6 are more appropriate than
models 4 and 7 which lack the high densities that appear to
be required. Models 5 and 6 are not fundamentally different. They
consist of a rather low density ($n(\h2) = 100 \cc$), hot ($T_{\rm kin}$ 
= 100--150 K) and very optically thick ($\tau$ = 15--20 for $J$=1--0)
component and a more dense, cooler component of relatively low optical 
thickness. The actual temperature and density of the latter component is 
not firmly established. As can be seen in Table~5, temperatures between 
20 and 60 K and densities between $10^{4}$ and 10$^{5} \cc$ can be 
obtained by varying the relative contributions of the two gas components, 
requiring only that the ratio $T_{\rm kin}^{2}/n(\h2)$ is kept constant.
The ionized carbon intensities can only be explained by postulating the 
presence of a hot and dense gas phase not sampled by the CO observations 
but rather resulting from extensively photodissociated and ionized gas
(interface with HII regions and supernova remnants?). A similar situation 
was found to apply in M~83 (Israel $\&$ Baas, 2001).

In any case, the molecular medium in the center of Maffei~2 appears to 
be significantly different from that in the center of IC~342. In Maffei~2,
the dense phase is warmer and probably denser, and its hot phase is
much less dense. Maffei~2 is experiencing a strong central starburst 
comparable to the one in IC~342. A difference between the two, at least 
at radio wavelengths, is that Maffei~2 and M~83 have much larger amounts 
of nonthermal emission than IC~342 (Turner $\&$ Ho 1983, 1994). This
suggests that the starbursts in Maffei~2 and M~83 are more evolved than 
the one in IC~342, perhaps explaining the molecular gas differences as well.

As the distance and metallicity of Maffei~2 have not been measured,
we have had to assume plausible values. This is an additional source
of uncertainty in the $\h2$ column densities, the masses and the
mass densities derived for Maffei~2 in Table~6. As the observed
HI column densities are much less than the model-inferred total H 
column densities, $\h2$ {\it column densities, masses and mass 
densities} scale linearly with the actual carbon abundance, whereas the 
$\h2$ and gas {\it masses} scale with the actual distance squared.
Because it is unlikely that the distance of Maffei~2 differs much from
the adopted value of 2.7 Mpc, the error in the derived mass due to this
uncertainty is not more than about a factor of two. The derived $X$ value is  
subject to uncertainties in the carbon abundance but not to those in the 
distance. The $N(\h2)$ values listed in Table~6 imply for Maffei~2 an $\h2$/CO 
conversion ratio $X = 2-3 \times 10^{19}$ $\h2$ mol cm$^{-2}$/$\kkms$, 
so that the same comment made for IC~342 also applies to Maffei~2.

\section{Conclusions}

\begin{enumerate}

\item Maps of the central arcminute of the nearby starburst galaxies IC~342
and Maffei~2 in various transitions of $\co$ and $\13co$,
and in [CI] confirm the compact nature of the central molecular gas 
emission in both galaxies. Most of this gas is within a 200 parsec 
from the nucleus. In both galaxies, the molecular gas seems to reside
in bright inner spiral arms, rather than a disk or torus.

\item Relative $^{12}$CO line intensities observed in matched beams are 
virtually identical in IC~342 and Maffei~2. However, relative $^{13}$CO
intensities, as well as the intensities of the [CI] and [CII]  lines
differ significantly, indicating different physical conditions for
the molecular gas in the two galaxies. The observed line ratios require 
modelling with a multi-component molecular medium in both galaxies. 

\item In IC~342, a dense component with $n(\h2) \approx  10^{4} \cc$ 
and $T_{kin} \approx$ 10 K is present together with a less dense 
$n(\h2) \approx 3 \times 10^{3} \cc$ and hotter $T_{kin} \approx$ 
100 K component. Total carbon column densities are about 1.5 times 
the CO column density. The modelling solution for IC~342 is reasonably
well-defined.

\item In Maffei~2, the parameters of the two components minimally 
required are less clearly defined. The more tenuous component has 
$n(\h2) \approx 10^{2} \cc$ and $T_{\rm kin} \approx$ 100 - 150 K, 
whereas the denser component has $n(\h2) \approx 10^{4} - 10^{5} \cc$ 
and $T_{\rm kin}$ = 20 -- 60 K. Maffei~2 seems to be more affected by 
CO dissociation as it has a C/CO ratio of two to three. 

\item In both starburst centers a significant fraction of the molecular 
mass (about half to two thirds) is associated with the hot PDR phase. 

\item With an estimated gas-phase [C]/[H] abundance of order 4 $\times 
10^{-4}$, the centers of NGC~6946 and M~83 contain within $R$ = 0.25 kpc 
similar (atomic and molecular) gas masses of about $10^{7}\, \Msun$. 
Peak face-on gas mass densities are typically 70 $\Msun$\, pc$^{-2}$ for 
IC~342 and 35 $\Msun$\, pc$^{-2}$ for Maffei~2, but the results for 
the latter are subject to relatively large errors caused by uncertainties 
in its distance and its carbon abundance.

\end{enumerate}

\acknowledgements

We are indebted to Ewine van Dishoeck and David Jansen for providing us 
with their detailed radiative transfer models. We thank the JCMT personnel 
for their support and help in obtaining the observations discussed in this 
paper. 


\begin{thebibliography}{}
%
\bibitem{} Buta R.J. $\&$ McCall M.L., 1999, \apjs 124, 33
\bibitem{} Crawford M.K., Genzel R., Townes C.H. $\&$ Watson D.M., 1985 
	\aua 291, 755
\bibitem{} Dahmen G., H\"uttemeister S., Wilson T.L. $\&$ Mauersberger R.,
	1998 \aua 331, 959
\bibitem{} Downes D., Radford S.J., Guilloteau S.,\, et al., 1992, \aua 
        262, 424
\bibitem{} Dressel L.L., Condon J.J. 1976, \apjs 31, 187
\bibitem{} Eckart A., Downes D., Genzel R., et al., 1990, \apj 348, 434
\bibitem{} Garnett D.R., Shields G.A., Skillman E.D., Sagan S.P. $\&$ 
        Dufour R.J., 1997, \apj 489, 63
\bibitem{} Garnett D.R., Shields G.A., Peimbert M.,\, et al. 1999 \apj 513, 168
\bibitem{} G\"usten R., Serabyn E., Kasemann C.,\, et al., 1993 \apj 402, 537
\bibitem{} Harris A.I., Hills R.E., Stutzki J.,\, et al. 1991. \apjl 382, L75
\bibitem{} Henkel C., Mauersberger R. $\&$ Schilke P., 1988, \aua 201, L23 
\bibitem{} Henkel C., Chin Y.-N, Mauersberger R. $\&$ Whiteoak J.B., 1998 
        \aua 329, 443
\bibitem{} Henkel C., Mauersberger R., Peck A.B., Falcke H. $\&$ Hagiwara Y.,
	2000, \aua 361, L45
\bibitem{} Ho P.T.P., Turner J.L. $\&$ Martin R.N., 1987, \apjl 322, L67
\bibitem{} Ho P.T.P., Martin R.N., Turner J.L. $\&$, Jackson J.M., 1990, \apjl 
	355, L19
\bibitem{} Huchtmeier W.K., Karachentsev I.D. $\&$ Karachentseva V.E., 2000 in:
	{\it Small Galaxy Groups}, IAU Symp. 174; ASP Conference Series Vol.
	209, Eds. M.J. Valtonen $\&$ C. Flynn; p. 158
\bibitem{} Hummel E. $\&$ Gr\"ave R., 1990 \aua 228, 315
\bibitem{} Hurt R.L. $\&$ Turner J.L., 1991, \apj 377, 434
\bibitem{} Hurt R.L., Turner J.L., Ho, P.T.P. $\&$ Martin R.N., 1993, 
        \apj 404, 602
\bibitem{} Hurt R.L., Turner J.L. $\&$ Ho P.T.P, 1996 \apj 466, 135
\bibitem{} H\"uttemeister S., Henkel C., Mauersberger R.,\, et al., 1995, 
	\aua 295, 571
\bibitem{} H\"uttemeister S., Mauersberger R. $\&$ Henkel C., 1997, 
        \aua 326, 59
\bibitem{} Irwin J.A. $\&$ Avery L.W., 1992, \apj 388, 328
\bibitem{} Ishiguro M., Kawabe R., Morita K.-I.,\, et al., 1989, \apj 344, 763
\bibitem{} Ishizuki S., Kawabe R., Ishiguro M., et al., 1990 Nature 344, 224
\bibitem{} Israel F.P., White G.J. $\&$ Baas F., 1995, \aua 302, 343
\bibitem{} Israel F.P. $\&$ Baas F., 1999 \aua 351, 10
\bibitem{} Israel F.P. $\&$ Baas F., 2001 \aua 371, 433
\bibitem{} Jackson J.M., Paglione T.A.D., Carlstrom J.E. $\&$ Rieu N.-Q., 
	1995, \apj 438, 695
\bibitem{} Jansen D.J., 1995, Ph.D. thesis, University of Leiden (NL)
\bibitem{} Jansen D.J., van Dishoeck E.F. $\&$ Black J.H., 1994, \aua, 282, 605
\bibitem{} Kaufman M.J., Wolfire M.G., Hollenbach D.J. $\&$ Luhman M.L., 1999
	\apj 527, 795
\bibitem{} Madore B.F. $\&$ Freedman W.L., 1992 \pasp 104, 362
\bibitem{} Martin R.N., Ho P.T.P. $\&$ Ruf K., 1982 Nature 296, 632
\bibitem{} Martin R.N. $\&$ Ho P.T.P., 1985 \apjl 308, L7
\bibitem{} Mauersberger R. $\&$ Henkel C., 1989 \aua 223, 79
\bibitem{} Mauersberger R. $\&$ Henkel C., 1991 \aua 245, 457
\bibitem{} Mauersberger R. $\&$ Henkel C., 1993 Rev. Mod. Astron. 6, 69
\bibitem{} Mauersberger R., Henkel C. $\&$ Chin Y.N., 1995, 294, 23
\bibitem{} Mauersberger R., Henkel C., Wilson T.L. $\&$ Harju J., 1989 
           \aua 226, L5
\bibitem{} Mauersberger R., Henkel C., Walsh W. $\&$ Schulz A., 1999 
           \aua 341, 256
\bibitem{} McCall M.L., 1989, \aj 97, 1341
\bibitem{} Meier D.S., Turner J.L. $\&$ Hurt R.L., 2000, \apj 531, 200
\bibitem{} Meier D.S. $\&$ Turner J.L., 2001 \apj 551, 687
\bibitem{} Morris M. $\&$ Lo K.Y., 1978 \apj 223, 803
\bibitem{} Newton K., 1980, \mnras 191, 169
\bibitem{} Oka T., Hasegawa T., Hayashi M., Handa T. $\&$ Sakamoto S., 1998
        \apj 493, 370
\bibitem{} Paglione T.A.D., Jackson J.M., Ishizuki S. $\&$ Rieu, N.-Q., 1995
	\aj 109, 1716
\bibitem{} Paglione T.A.D., Jackson J.M. $\&$ Ishizuki S., 1997 \apj 484, 656
\bibitem{} Petit-Pas G.R. $\&$ Wilson C.D., 1998 \apj 503, 219
\bibitem{} Rickard L.J, Palmer P., Morris M., Turner B.E. $\&$ Zuckerman B., 
	1977 \apj 213, 673
\bibitem{} Rickard L.J $\&$ Blitz L., 1985 \apjl 292, L57
\bibitem{} Rickard L.J $\&$ Palmer P.M., 1981 \aua 102, L13
\bibitem{} Rieu, N.-Q., Henkel C., Jackson J.M. $\&$ Mauersberger R., 
	1991, \aua 241, L33
\bibitem{} Rieu, N.-Q., Jackson J.M., Henkel C., Bach T. $\&$ Mauersberger R., 
	1992, \apj 399, 521
\bibitem{} Rots A.H., 1979 \aua 80, 255
\bibitem{} Sage L.J., Shore S.N. $\&$ Solomon P.M., 1990, \apj 351, 422
\bibitem{} Sage L.J. $\&$ Isbell D.W., 1991 \aua 247, 320
\bibitem{} Sakamoto K., Okumura S.K., Ishizuki S. $\&$ Scoville N.Z., 1999 
	\apjs 124, 403
\bibitem{} Sandage A. $\&$ Tammann G.A., 1987, {\it A Revised Shapley-Ames
	Catalog of Bright Galaxies}, second edition, Carnegie Institution of
	Washington Publication 635 (Washington, D.C.: Carnegie Institution of
	Washington).
\bibitem{} Sargent A.I., Sutton E.C., Masson C.R., Lo, K.Y. $\&$ Phillips T.G.,
	1985, \apj 289, 150
\bibitem{} Sodroski T.J., Odegard N., Dwek E.,\, et al. 1995 \apj 452, 262
\bibitem{} Stacey G.J., Geis N., Genzel R.,\, et al. 1985 \aua 373, 423
\bibitem{} Steppe H., Mauersberger R., Schulz A. $\&$ Baars J.W.M., 1990, 
        \aua 233, 410
\bibitem{} Stutzki J., Graf U.U., Honingh C.E.,\, et al. 1997, \apjl 477, 33
\bibitem{} Takano S., Nakai N., Kawaguchi K. $\&$ Takano T., 2000, 
        \pasj, 52, L67
\bibitem{} Turner J.L. $\&$ Ho P.T.P., 1983 \apj 268, L79
\bibitem{} Turner J.L. $\&$ Ho P.T.P., 1994 \apj 421, 122
\bibitem{} Turner J.L. $\&$ Hurt R.L., 1992, \apj 384, 72
\bibitem{} Turner J.L., Hurt R.L. $\&$ Hudson D.Y., 1993, \apjl 413, L19
\bibitem{} van Dishoeck E.F. $\&$ Black J.H., 1988, ApJ 334, 771
\bibitem{} Vila-Costas M.B. $\&$ Edmunds M.G. 1992 \mnras 259, 121
\bibitem{} Wall W.F. $\&$ Jaffe D.T., 1990, \apjl 361, L45
\bibitem{} Wall W.F., Jaffe D.T., Israel F.P., Bash F.N., Maloney P.R.
        $\&$ Baas F., 1993 \apj 414, 98 
\bibitem{} Weliachew L., Casoli F. $\&$, Combes F., 1988, \aua 199, 29
\bibitem{} White G.J., Ellison B., Claude S., Dent W.R.F. $\&$ Matheson D.N.,
	1994, \aua 284, L23
\bibitem{} Wright M.C.H., Ishizuki S., Turner J.L., Ho P.T.P $\&$ Lo K.Y.,  
        1993 \apj 406, 470
\bibitem{} Xie S., Young J.S. $\&$ Schloerb F.P., 1994, \apj 421, 434
\bibitem{} Young J.S. $\&$ Sanders D.B., \apj 302, 680
\bibitem{} Young J.S. $\&$ Scoville N.Z., 1982 \apj 258, 467
\bibitem{} Zaritsky D., Kennicutt R.C. $\&$ Huchra J.P., 1994, \apj 420, 87
%
\end{thebibliography}
\end{document}